 \documentclass[journal,twoside]{IEEEtran}

 	\usepackage[pdftex]{graphicx}
	\graphicspath{{../pdf/}{../jpeg/}}
	\DeclareGraphicsExtensions{.pdf,.jpeg,.png}    

	\usepackage[cmex10]{amsmath}
	\usepackage{mathabx}
	\usepackage{algorithmic}
	\usepackage{array,multirow}
	\usepackage{mdwmath}

      \usepackage{arydshln}                
      \RequirePackage{doi}

   \usepackage{hyperref}
   \usepackage[normalem]{ulem}
         \usepackage{caption}
        \usepackage{subcaption}
\usepackage{eqparbox}
    \newcolumntype{P}[1]{>{\centering\arraybackslash}p{#1}} 

\newlength{\Oldarrayrulewidth}
\newcommand{\Cline}[2]{%
  \noalign{\global\setlength{\Oldarrayrulewidth}{\arrayrulewidth}}%
  \noalign{\global\setlength{\arrayrulewidth}{#1}}\cline{#2}%
  \noalign{\global\setlength{\arrayrulewidth}{\Oldarrayrulewidth}}}

 \usepackage{authblk}  
\begin{document}

\title{Effect of  Edge Roughness on resistance and switching voltage of Magnetic Tunnel Junctions}


\renewcommand*{\Authands}{, }
\author[1]{Rachit R. Pandey}
\author[1]{Sutapa Dutta}
\author[1]{Heston A. Mendonca}

\author[1]{ Ashwin A. Tulapurkar}

\affil[1]{Solid State devices group, Department of Electrical Engineering, Indian Institute of Technology Bombay, Mumbai 400076,India}


\maketitle

\begin{abstract}

 We investigate the impact of edge roughness on the electrical transport properties of magnetic tunnel junctions using non-equilibrium Green’s function formalism. We have modeled edge roughness as a stochastic variation in the cross-sectional profile of magnetic tunnel junction characterized by the stretched exponential decay of the correlation function. The stochastic variation in the shape and size changes the transverse energy mode profile and gives rise to the variations in the resistance and switching voltage of the magnetic tunnel junction. We find that the variations are larger as the magnetic tunnel junction size is scaled down due to the quantum confinement effect. 
A model is proposed for the efficient calculation of edge roughness effects by approximating the cross-sectional geometry to a circle with the same cross-sectional area. Further improvement can be obtained by approximating the cross-sectional area
to an ellipse with an aspect ratio determined by the first transverse eigenvalue corresponding to the 2D cross section. These results would be useful for reliable design of the spin transfer torque- magnetic random access memory (STT-MRAM) with ultra-small magnetic tunnel junctions.
\end{abstract}

\IEEEoverridecommandlockouts

\IEEEpeerreviewmaketitle


\section{Introduction}
Magnetic tunnel junction (MTJ) comprises two ferromagnetic layers (free layer and pinned layer) separated by a tunneling barrier. Binary information can be stored in MTJs corresponding to parallel (P) and anti-parallel (AP) configurations of the magnetizations. The information can be read by measuring the resistance which is low for P and high for AP configurations respectively.  Spin transfer torque (STT) produced by application of large positive and negative voltages to free layer with respect to the fixed layer, stabilizes P and AP configurations respectively, and thus can be used for writing the memory.  As ferromagnetic layers with perpendicular magnetic anisotropy (PMA) have lower threshold switching voltage ($V_c$) with enhanced thermal stability, they are preferred over in-plane magnetized layers \cite{Ikeda_PMA}. Reliability analysis of STT-MRAM (Magnetic Random Access Memory) in terms of write error, tunnel oxide breakdown, temperature variations, etc. has been carried out before \cite{IBM_2016, Chappert_2012, PVT_variation, Zhang, Samsung, zhao_2016}. 
In this paper, we have investigated the effect of lithographic imperfections on the performance of MTJ, which becomes more evident with the technology scaling.  The circular edge roughness (CER) is defined as the straying of a pattern from its expected circular shape and is used to characterize the unwanted sidewall roughness emerging during fabrication processes \cite{ Ban_2010, Amita_2017, ReRAM_2019}. The threshold voltages and resistances of the MTJ have been calculated using non-equilibrium Green's function (NEGF) method, for 250 realizations of the sidewalls for fixed CER parameters. CER affects the area as well as the shape of MTJ, which in turn changes the transverse mode energies of the electrons tunneling across the barrier and thus gives rise to variations in the resistance and threshold voltage. We have calculated these variations for a range of CER parameters. The NEGF calculation needs transverse energy eigenvalues which were obtained by solving the Schrodinger equation for a 2d potential well with a random boundary corresponding to each realization.




\section{Simulation Methodology}

\begin{figure}[ht!] 
\centering
\includegraphics[width= 3.4 in,keepaspectratio]{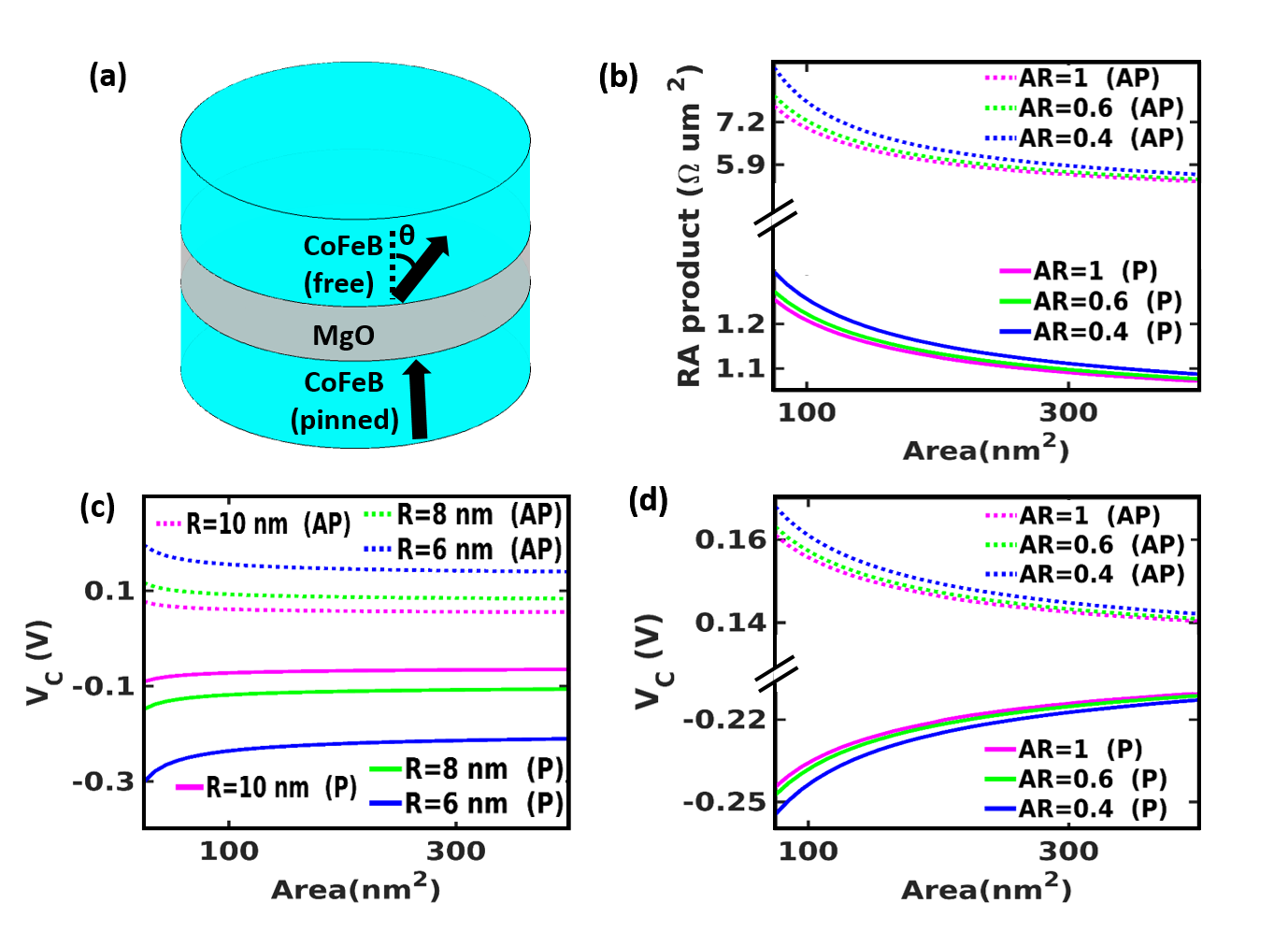}
\caption{(a) Schematic of MTJ without edge roughness. (b) RA product vs area for ellipse with different aspect ratios (AR=1 corresponds to a circle).(c) $V_c$ vs area for circle with energy barrier of $40 k_B T$ for radii=6 nm (blue), 8nm (green) and 10 nm (magenta). (d) $V_c$ vs area for ellipse with different aspect ratios for energy barrier of $40 k_B T$ for radius=6 nm.}
\label{smooth_mtj}
\end{figure}

In the first step, charge current and spin current for a given applied voltage across MTJ is calculated using 1d NEGF formalism, as a function of transverse mode energy at 300 K temperature.  The device Hamiltonian matrix is modeled using an effective mass tight binding approach. The transport of electrons across the device is assumed to be coherent.  The effect of the contacts is taken into account as self-energy contributions to the Hamiltonian. Charge and spin currents are calculated from the energy-resolved electron correlation matrix \cite{datta_2005, Datta_2009}.
We used CoFeB as the ferromagnet for both the fixed and free layers with the Fermi energy $E_F$ =2.25 eV and the exchange splitting, $\Delta$  = 2.15 eV. The barrier height from the Fermi level is taken as $U_B=$ 0.76 eV. The effective mass of MgO (tunnelling barrier) and FM are taken as 0.16 $m_e$ and 0.38 $m_e$, respectively, where $m_e$ is the free electron mass. The thickness of the oxide layer is set to 0.9 nm. The charge and parallel spin current (spin current along the fixed layer direction) as a function of the transverse mode energy are tabulated for a range of voltage values ranging from -0.6 to 0.6 V for both P and AP configurations.  
In the second step, the transverse energy modes are found from the solution of the Schrodinger equation for 2d infinite well with a boundary given by the cross-section of the MTJ. If the cross-section is a perfect circle, the eigenvalues of Hamiltonian are known analytically. For an arbitrary cross-section, the eigenvalues can be found numerically using finite difference method by discretizing the area into a square grid.
In the third step, the charge current and spin current for each transverse mode are summed up to get the net charge and spin current for a range of voltage values ranging from -0.6 to 0.6 V for both P and AP configurations.
The resistance-area (RA) product calculated at 0.01 V, for MTJ with elliptical cross-section for different aspect ratios as a function of corresponding areas is shown in Fig. \ref{smooth_mtj}b. From this figure, we can see that as the area reduces, the RA product shows  dependence on area as well as shape \cite{Debasis_2018}.  
The critical spin current can be calculated from the Gilbert damping ($\alpha_G$) and the energy barrier between P and AP states ($\Delta E$), as
$I_{sc}=(4q\alpha_G/\hbar) \Delta E$. Further, the energy barrier is given by, $\Delta E=(1/2) \mu_0 M_s A 
 t_{FM} H_K $, where  $M_s$, $A$, $t_{FM}$, $H_{K}$  denote the saturation magnetization, cross-sectional area, free layer thickness and effective perpendicular anisotropy respectively. 
The critical voltage can be found by interpolating spin current vs voltage data. If the radius of MTJ is 10 nm,  assuming $\Delta E=40 k_B T$ (T=300 K), $\alpha_G$=0.08, $t_{FM}$=2 nm and $M_s=1.2 \times 10^6 A/m$, the $H_{K}$ comes out to be $3.5\times 10^5 A/m$.  The critical voltage for P to AP and AP to P  switching as a function of area assuming circular cross-section and the same $H_{K}$ is shown by the magenta curve in Fig. \ref{smooth_mtj}c. Similar calculations for 8 nm and 6 nm radii are shown by green and blue curves respectively. Fig. \ref{smooth_mtj}d shows the critical voltage  (assuming $H_{K}$ of 6 nm radius MTJ) for elliptical cross-section of different aspect ratios as a function of the area.
We can see that as the area reduces, the threshold voltage shows  dependence on area as well as shape. 


\begin{figure}[ht!] 
\centering
\includegraphics[width= 3.4 in,keepaspectratio]{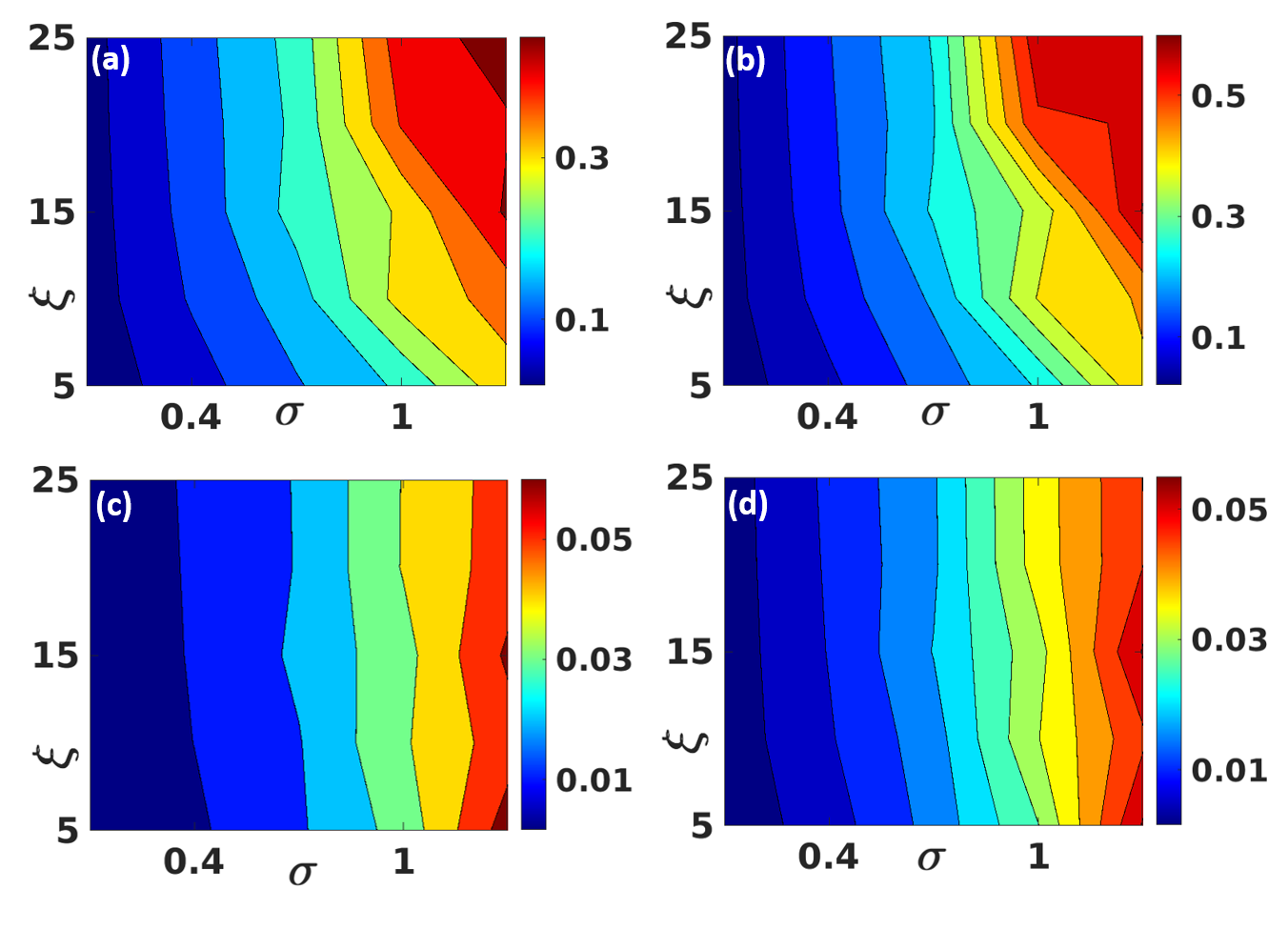}

\caption{  Coefficient of Variation plots for: (a)  Resistance (P) (b) Resistance (AP) (c) $V_c$ (P) (d) $V_c$ (AP). }
\label{2d_plots}
\end{figure}


\begin{figure}[ht!] 
\centering
\includegraphics[width= 3.4 in,keepaspectratio]{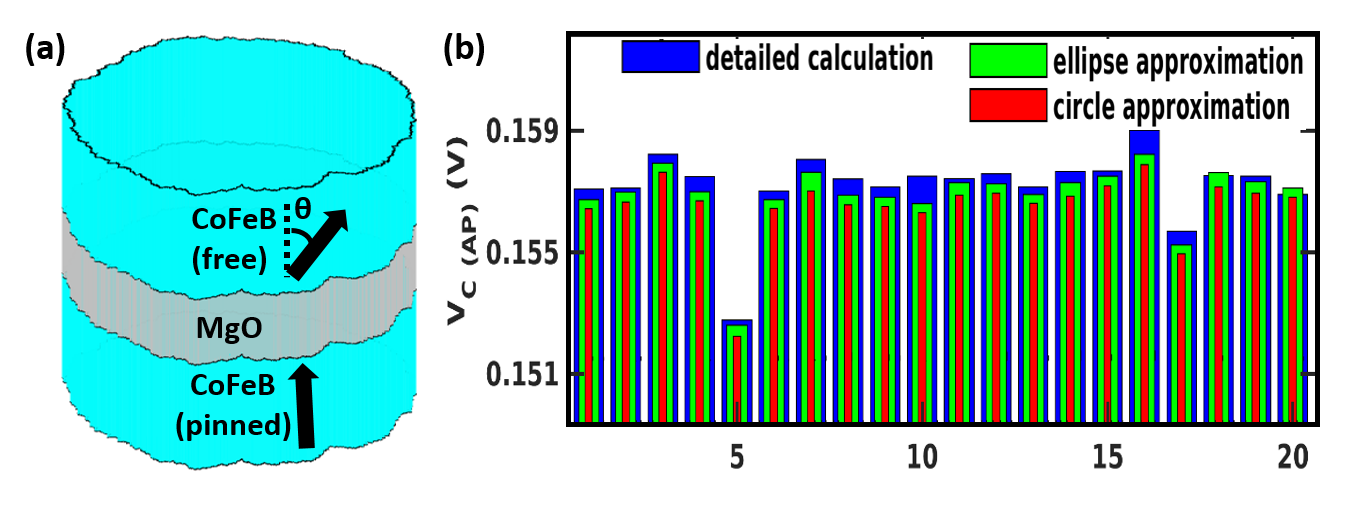}

\caption{ (a) Schematic of MTJ with edge roughness. (b) comparison of  $V_c$ (AP)  for 20 trials obtained from detailed calculation(blue), circle approximation (red), ellipse approximation  (green).}
\label{bar_plots}
\end{figure}

For incorporation of circular edge roughness into a circular cross-section of radius $R_0$, we make a random line segment of length $2 \pi R_0$ with auto-correlation function (R) given by the  equation, $R(x)=\sigma^2e^{-(d/\xi)^{2\alpha}}$,
where the chord length $d$ is given by $d=2 R_0 |sin(x/2R_0)|$. $\xi$, $\alpha$, and $\sigma$ denote the correlation length, roughness parameter and standard deviation respectively \cite{bio_SOUBEYRAND, Gogolides_2011} . A realization of random line segment is obtained as follows \cite{zhao2000characterization}: We numerically generate white noise series with unit power spectral density (PSD) and take its Fourier transform. This is then multiplied by the PSD of the correlation function. The inverse FT of the product gives us a random line segment. The random shape is constructed by taking $R_0+x$ as the radii distribution for angles from 0 to 2$\pi$.
The coefficient of variation (CV=standard deviation/mean) for the quantities to be analyzed is obtained from 250 samples.


\section{Results and Discussions}

Variation in the area and shape of MTJ cross-section due to the CER produces variation in the transverse energy mode profile. This in turn produces variation in the charge current and spin current flowing across the MTJ for a given applied voltage. The coefficient of variation of resistance and switching voltage as a function of $\sigma$ and $\xi$ for $\alpha=0.5$ and average radius 6 nm obtained from detailed calculation is shown as a 2D plot in Fig. \ref{2d_plots}. We can see that the variations become larger as $\sigma$ and $\xi$ increase. CV for different parameters at the centre of 2D plot ($\sigma=0.67 nm$, $\xi=15 nm$) are shown in the table \ref{table1} for different average radii of the cross-section under ``detailed calculation" column heading. We can see that the variations increase as the MTJ size is scaled down. To find out the influence of area variation, for each of the 250 samples, we mapped the random shape to a perfect circle of the same area and found out the resistance and switching voltage (See Fig. \ref{smooth_mtj}c). The CV obtained from this procedure is shown in table \ref{table1} under ``circle approximation" coumn heading and it matches well with values obtained from detailed calculation. 

The circle approximation is expected to work well when the ratio, ($\sigma/R_0$) is small. Further, for the approximation to work well, the minimum normalized correlation function $e^{-(2 R_0/\xi)^{2\alpha}}$ should be close to 1 i.e.$(2 R_0/\xi)^{2\alpha}$ should be small. 
If area variation due to CER plays a dominant role, we can  estimate the variance in a quantity Q as, 
\begin{equation}
var(Q) \approx (dQ/dA)^2 [2 \int_{0}^{L} (L-x)R(x) dx] 
\label{eqn1}
\end{equation}
where  $L=2\pi R_0$ is the average perimeter. The term in the square bracket in the above equation is the area variance.
The CV of various parameters estimated with above equation is given under ``estimated" column heading in table \ref{table1}. We can see that values estimated from area variation are fairly close to the numerically calculated values. 
These equations imply that the area variance is proportional to $\sigma^2$ and it is an increasing function of $\xi$,  which is consistent with trends seen in the 2d plots in Fig. \ref{2d_plots}. (area variance  saturates at large values of $\xi/L$).


\setlength\arrayrulewidth{1pt}
\begin{table}
\caption{$\%$ CV for $\alpha=0.5$ $\sigma=0.67nm$ $\xi=15nm$}
\normalsize
\centering
\begin{tabular}{ |P{0.55in}|P{0.5in}|P{0.7in}|P{0.7in}|P{0.5in}|  }
 \hline
$\%$  CV of & $R_0(nm)$ & Detailed calculation & Estimated from eq. \ref{eqn1} & Circle approx. \\
\hline 
                & 6 & 21.73 &20.5 & 21.62 \\        \Cline{.1pt}{2-5}
$R_{P}$         & 8 & 12.94 &13.86 & 12.95 \\       \Cline{.1pt}{2-5}
                & 10 & 10.12 &10.20 & 9.99 \\       
\hline 

                & 6 &  25.63 &23.32 & 25.33  \\     \Cline{.1pt}{2-5}
$R_{AP}$        & 8 &  14.26 &15.04 & 14.15  \\      \Cline{.1pt}{2-5}
                & 10 & 10.89 &10.93 & 10.69  \\

\hline 
                & 6 & 2.18& 2.07 & 1.99 \\          \Cline{.1pt}{2-5}
$Vc_{P}$        & 8 & 0.93& 0.92 & 0.94\\           \Cline{.1pt}{2-5}
                & 10 & 0.57& 0.57 & 0.56 \\          
\hline 
                & 6 & 2.02 &1.83 & 1.85  \\        \Cline{.1pt}{2-5}
 $Vc_{AP}$      & 8   & 0.90& 0.90 & 0.89  \\       \Cline{.1pt}{2-5}
                & 10  & 0.55& 0.53 & 0.54  \\       
\hline 
\end{tabular}
\label{table1}
\end{table}
To see if the circle approximation can be further improved, we mapped a given random shape to an ellipse. This is done as follows: We first note down the area. We calculate numerically the ground state energy of the 2d infinite well with boundary given by the random edge. We then compare ground state energy with the tabulated ground state energies of ellipses with the same area and different aspect ratios. An aspect ratio is assigned to the random figure by interpolation. Using tabulated data of $V_c$ and resistance as a function of area for different aspect ratios (see Fig. \ref{smooth_mtj}), we can calculate the switching voltage and resistance of the random cross-section MTJ by interpolation. Fig. \ref{bar_plots} b shows the $V_c$ for AP to P state for 20 different realizations (out of 250). The blue bar corresponds to $V_c$ calculated by numerically ``exact" way i.e. getting all the transverse energy modes to form the numerical solution of 2d Schrodinger equation and summing up transverse currents for each mode. The green bar corresponds to the calculation by mapping the shape to an ellipse which needs only the ground state energy calculation and is hence faster.
However, for large values of $\sigma/R_0$ and $R_0/\xi$, the contribution from the non-elliptical shape variation should be taken into account.
It should be also noted that the area variation arising from CER gives rise to variation in the thermal stability as the energy barrier $\Delta E$, depends on the area.     



\section{Conclusion}
We have demonstrated that edge roughness gives rise to variance in the area and shape of a magnetic tunnel junction. This in turn produces variance in the resistance and switching voltage. The variance becomes larger as the MTJ size is scaled down. These results would be useful for designing reliable MRAM cells.


\bibliographystyle{IEEEtranDOI}
\bibliography{references_file}

\end{document}